\documentclass[fleqn,usenatbib]{mnras}
\usepackage{newtxtext,newtxmath}

\usepackage[T1]{fontenc}
\usepackage{ae,aecompl}

\usepackage{graphicx}	
\usepackage{times}

\title[RSG atmospheres with winds]{The impact of winds on the spectral appearance of Red Supergiants}

\author[Davies \& Plez]{
Ben Davies$^{1}$\thanks{b.davies@ljmu.ac.uk}, and Bertrand Plez$^{2}$
\\
$^{1}$Astrophysics Research Institute, Liverpool John Moores 
University, Liverpool Science Park ic2, mv 146 Brownlow Hill, Liverpool, L3 5RF, UK.\\
$^{2}$Laboratoire Univers et Particules de Montpellier, Univ Montpellier, CNRS, Montpellier, France.\\
}

\date{Accepted XXX. Received YYY; in original form ZZZ}

\pubyear{2017}
\hypersetup{draft}
\begin{document}
\label{firstpage}
\pagerange{\pageref{firstpage}--\pageref{lastpage}}
\maketitle
      
\begin{abstract}
  The rate at which mass is lost during the Red Supergiant evolutionary stage may strongly influence how the star appears. Though there have been many studies discussing how RSGs appear in the mid and far infrared (IR) as a function of their mass-loss rate, to date there have been no such investigations at optical and near-IR wavelengths. In a preliminary study we construct model atmospheres for RSGs which include a wind, and use these models to compute synthetic spectra from the optical to the mid-infrared. The inclusion of a wind has two important effects. Firstly, higher mass-loss rates result in stronger absorption in the TiO bands, causing the star to appear as a later spectral type despite its effective temperature remaining constant. This explains the observed relation between spectral type, evolutionary stage and mid-IR excess, as well as the mismatch between temperatures derived from the optical and infrared. Secondly, the wind mimics many observed characteristics of a `MOLsphere', potentially providing an explanation for the  extended molecular zone inferred to exist around nearby RSGs. Thirdly, we show that wind fluctuations can explain the spectral variability of Betelgeuse during its recent dimming, without the need for dust. 
\end{abstract}


\def\ga{\mathrel{\hbox{\rlap{\hbox{\lower4pt\hbox{$\sim$}}}\hbox{$>$}}}}
\def\la{\mathrel{\hbox{\rlap{\hbox{\lower4pt\hbox{$\sim$}}}\hbox{$<$}}}}
\def\msunyr{M\mbox{$_{\normalsize\odot}$}\rm{yr}$^{-1}$}
\def\msun{$M$\mbox{$_{\normalsize\odot}$}}
\def\zsun{$Z$\mbox{$_{\normalsize\odot}$}}
\def\rsun{$R$\mbox{$_{\normalsize\odot}$}}
\def\rstar{$R_\star$}
\def\minit{$M_{\rm init}$}
\def\lsun{$L$\mbox{$_{\normalsize\odot}$}}
\def\mdot{$\dot{M}$}
\def\logmdot{$\log(\dot{M}/{\rm M_\odot\,yr^{-1}})$}
\def\mdotdj{$\dot{M}_{\rm dJ}$}
\def\lbol{$L$\mbox{$_{\rm bol}$}}
\def\kms{\,km~s$^{-1}$}
\def\EW{$W_{\lambda}$}
\def\arcsec{$^{\prime \prime}$}
\def\arcmin{$^{\prime}$}
\def\teff{$T_{\rm eff}$}
\def\Teff{$T_{\rm eff}$}
\def\logg{$\log g$}
\def\logz{$\log Z$}
\def\logl{$\log (L/L_\odot)$}
\def\vdisp{$v_{\rm disp}$}
\def\vinf{$v_{\infty}$}
\def\bcv{{\it BC$_V$}}
\def\bci{{\it BC$_I$}}
\def\bck{{\it BC$_K$}}
\def\lmax{$L_{\rm max}$}
\def\um{$\mu$m}
\def\chisq{$\chi^{2}$}
\def\AV{$A_{V}$}
\def\hminus{H$^{-}$}
\def\Hminus{H$^{-}$}
\def\ebmv{$E(B-V)$}
\def\mdyn{$M_{\rm dyn}$}
\def\mphot{$M_{\rm phot}$}
\def\cnterm{[C/N]$_{\rm term}$}
\newcommand{\fig}[1]{Fig.\ \ref{#1}}
\newcommand{\Fig}[1]{Figure \ref{#1}}
\newcommand{\newtext}[1]{{#1}}
\newcommand{\nntext}[1]{{#1}}


\section{Introduction}
It is well-established that mass-loss plays a pivotal role in the evolution of massive stars, impacting the path across the Hertzsprung-Russell (H-R) diagram, the time spent in the various evolutionary phases, and the appearance of the supernova \citep{Chiosi-Maeder86}. Of the total mass lost, a substantial fraction of this may be during the post main-sequence (MS), for example when the star is a Red Supergiant (RSG). Though this phase lasts only $\la$ one tenth of the main sequence, the star may lose several solar masses of envelope during this stage of evolution \citep[][]{Smith14,Beasor20}.

Much of what we know about mass-loss during the {\it hot} phases of massive star evolution comes from spectral synthesis. The transfer of radiation is computed through an expanding atmosphere and the resulting spectral appearance determined, which is then compared to observations \citep[e.g.][]{Kudritzki-Puls00,Hillier03,Markova18}. By contrast, there have been surprisingly few studies of how winds alter the spectra of cool massive stars. Studies that do exist tend to focus on the dusty region of the wind, observable as continuum emission in the mid-infrared (IR) \citep[e.g.][]{vanLoon99,Groenewegen09,Beasor-Davies18}, or molecular line emission in the far-IR \citep{decin06,debeck10}. Though ultraviolet studies exist, these tend to focus on modelling individual objects rather than being predictive \citep[see e.g. the study of $\gamma$~Vel in][]{Carpenter99}. To date, there has been very little attention paid to how winds impact the optical to near-infrared region of RSGs as a function of mass-loss rate. 

One way in which one might expect a wind to augment the optical spectral appearance of a RSG is through enhanced TiO absorption. The TiO bands cover large parts of the optical spectral energy distribution, and are the most prominent spectral feature in RSGs. At fixed chemical abundances, the TiO molecule becomes more abundant at lower temperatures ($<$4000K), and hence contributes more to the opacity. Adding an expanding wind to a static model atmosphere of a RSG would increase the density in the outer layers where the gas is cooler, leading to increased opacity from the TiO bands. Therefore, the effect of a high mass-loss rate \mdot\ may be enhanced TiO absorption, which in turn would cause the star to appear as a later spectral type. If true, this would go some way to explaining several observed phenomenon such as the disagreement between measurements of effective temperatures in the optical and infrared \citep{Levesque05,rsgteff}, and the correlation of mass-loss rate with spectral type (see Section \ref{sec:spt-mdot}). 

In this paper we present an initial study in which we take the first steps towards spectral synthesis of RSG winds. Though we make a number of simplifying assumptions, we will demonstrate that stronger winds lead to later spectral types for a given underlying stellar effective temperature. The paper is organised as follows: in Sect.\ \ref{sec:method} we describe our simple wind model and our methodology for computing the radiative transfer. In Sect.\ \ref{sec:results} we present our results, and discuss their application to various general observations of RSGs in Sect.\ \ref{sec:disc}. We conclude in Sect.\ \ref{sec:conc} and include a description of our future development plans. 




\begin{figure}
\begin{center}
\includegraphics[width=8.5cm,bb=0 0 283 566,clip]{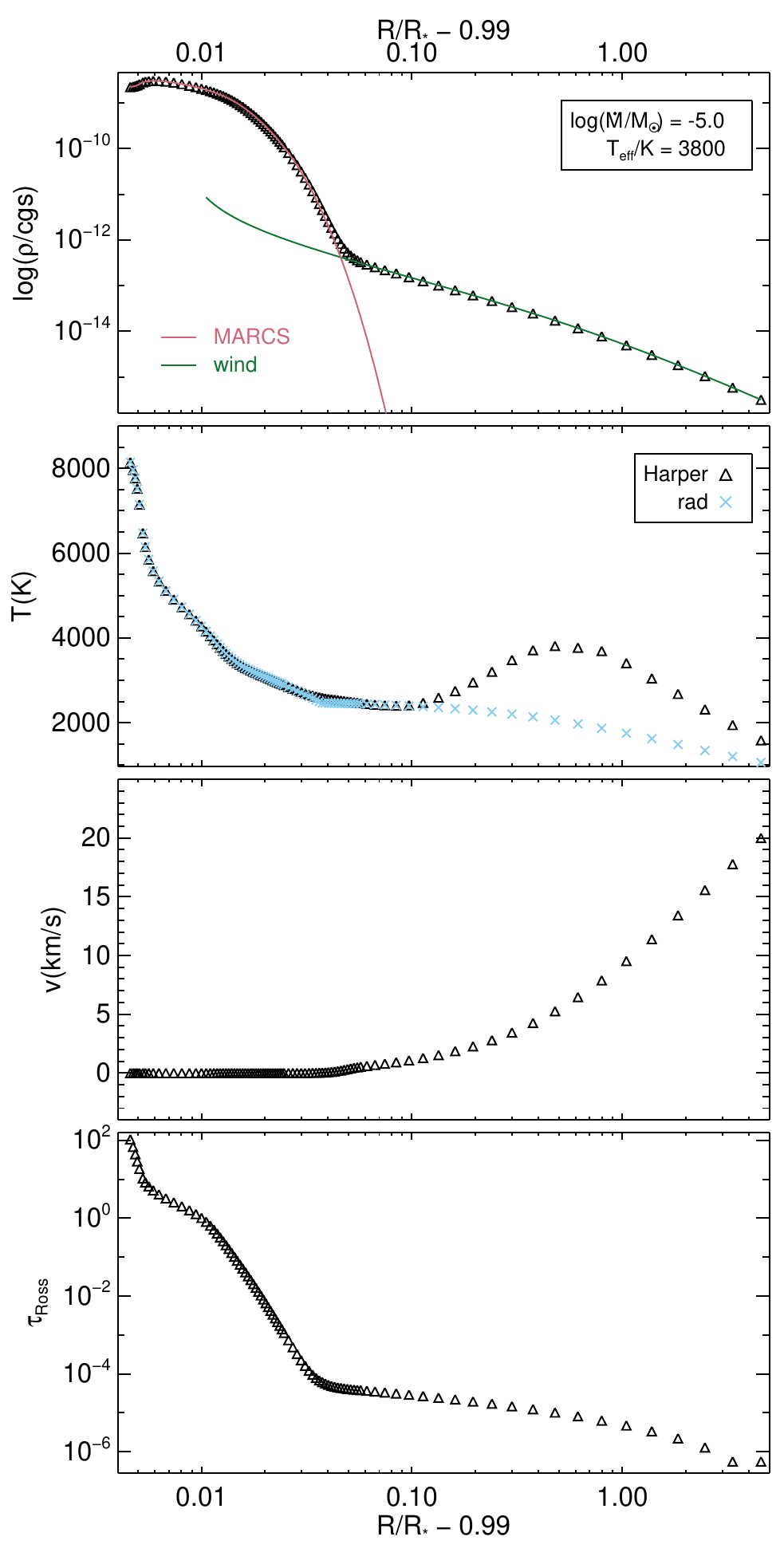}
\caption{Radial profiles of density (top), temperature (2nd top), velocity (2nd bottom) and Rosseland optical depth (bottom), of our $\log(\dot{M}/{\rm M_\odot\,yr^{-1}})$=-5 model. The temperature panel shows both the radiative equilibrium profile and the empirical profile from Harper et al.\ (2001). }
\label{fig:modprof}
\end{center}
\end{figure}

\section{Methodology} \label{sec:method}

\subsection{Model setup}
Firstly, we begin by creating a description of the state variables throughout the expanding atmosphere as a function of distance $r$ from the centre of the star. Our starting point is a MARCS model atmosphere \citep{gustafsson08} with effective temperature \teff=3800\,K, gravity \logg=0.0, Solar metallicity, microturbulent velocity 5\kms, luminosity $10^5$\lsun, and radius 730\rsun. To this, we add a wind of constant mass-loss rate \mdot\ in order to determine the density out to larger $r$ from mass continuity, 

\begin{equation}
\dot{M} = 4 \pi r^2 \rho(r) v(r)
\label{eq:masscont}
\end{equation}

\noindent where $\rho$ and $v$ are the density and velocity as a function of $r$ respectively. For the density profile $\rho_{\rm wind}(r)$, we use the model of $\alpha$~Ori constructed by \citet{Harper01} from spatially-resolved radio continuum data. The absolute scaling for $\rho_{\rm wind}(r)$ is determined by calculating the mass-loss rate at the outer boundary, then rescaling to the desired mass-loss rate. The velocity profile $v(r)$ is determined simply from Eq.\ (\ref{eq:masscont}) and a fiducial terminal wind speed \vinf, which we match to \citet{Harper01}. 


For the temperature profile $T(r)$, we again adopt the \citet{Harper01} model of $\alpha$~Ori. In this model, there is a chromospheric region above the classical photosphere in which the temperature gradient becomes positive with increasing $r$, peaking at around 1.4\rstar\ (see \fig{fig:modprof}), before declining again. The figure also shows how $T(r)$ departs from that expected of radiative equilibrium. The impact on our results of modifying the peak chromospheric temperature will be explored in Sect.\ \ref{sec:chromo}.


\subsubsection{Defining the grid of depth points}
When computing the spectra, we must integrate the source function from the outer wind to beneath the photosphere. To do this we must define the grid over which the source function is computed. For the photospheric part of the atmosphere, we simply use the MARCS depth points. For the wind region, we grid in optical depth. To do this we first determine a rough approximation of the run of Rosseland optical depth $\tau_{\rm Ross}$ throughout the atmosphere using MARCS opacity tables. We then resample the wind region of the atmosphere to be gridded in equal steps of $\Delta\log(\tau_{\rm Ross})=0.05$. The choice of  $d\log(\tau_{\rm Ross})$ was driven by two factors: when the grid is too coarse, the $\tau_\lambda=1$ surface is poorly resolved for strong absorption lines, leading to errors in the equivalent width of those lines. When the grid is too finely sampled, rounding errors can occur if the contribution of the source function from grid-point to grid-point is too small, leading to numerical difficulties. In resolution tests, we found that computed spectra were consistent when $\Delta\log(\tau_{\rm Ross})$ was set between 0.05 and 0.01.


The outer boundary of the grid is defined to be where the local temperature is 800K. Below these temperatures, our molecular equilibrium calculations are not stable, whilst in reality we might expect certain molecular species to begin depleting onto dust grains. This limiting temperature is reached at $\sim 10 R_\star$. The Rosseland optical depth at this radius is below $10^{-5}$ for even the highest \mdot\ models computed here. A wavelength-dependent depth analysis reveals that the $\tau_\lambda = 1$ surface is typically well within $10R_\star$. In our highest mass-loss rate models (\mdot $> 10^{-5}$\msunyr), we find that a small number of atomic resonance lines in the optical, e.g. Na\,{\sc i}, some atomic lines at $\lambda <0.4$\um, and some CO lines at $\sim$5\um\ are still optically thick at the outer edge of the grid.




\begin{figure*}
\begin{center}
\includegraphics[width=17cm]{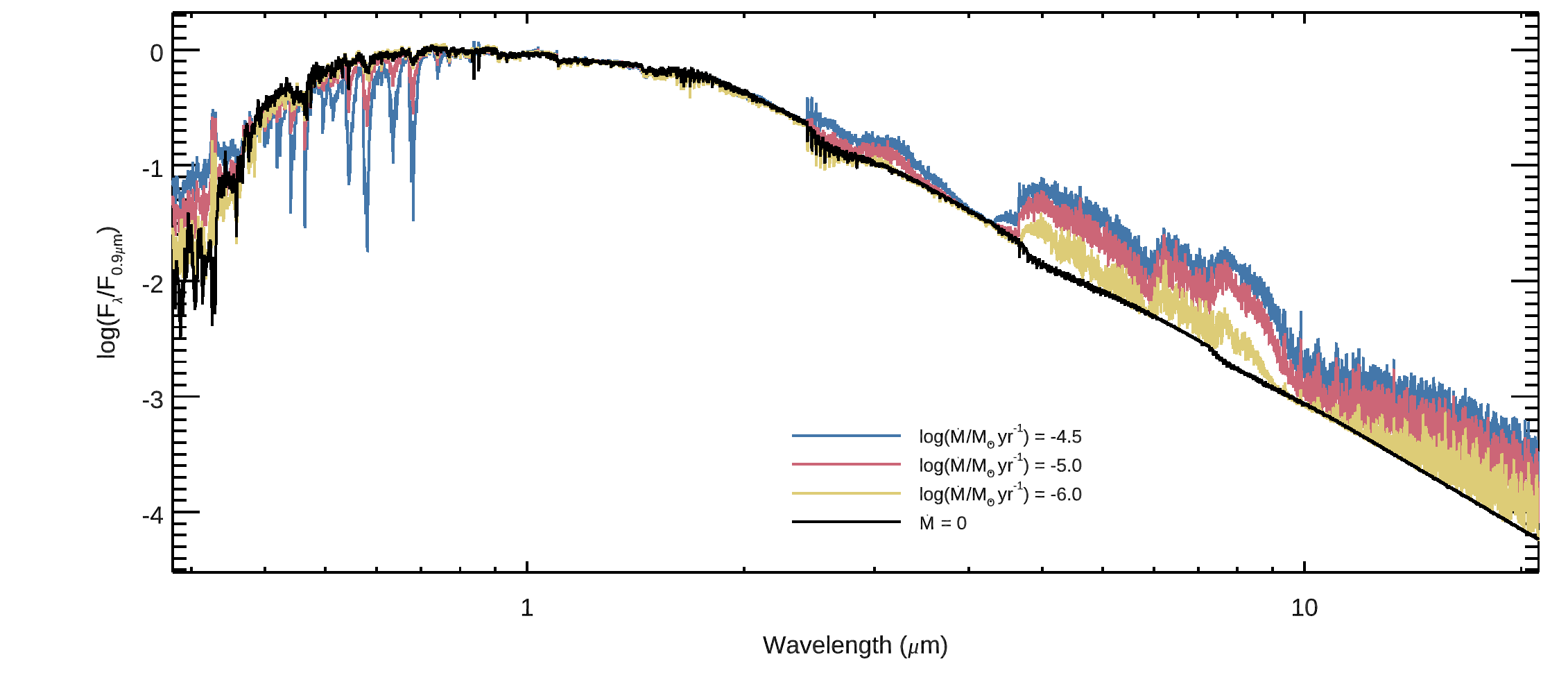}
\caption{Optical/IR spectra of our models. All have \teff=3800\,K, \logg=0.0, Solar metallicity and Solar-scaled relative abundances. The lines of different colours have the mass-loss rates indicated in the legend, in units of \logmdot. The spectra have been smoothed down to a resolving power $R=1000$ for clarity. Spectra are available for download via CDS .....}
\label{fig:spectrum}
\includegraphics[width=17cm]{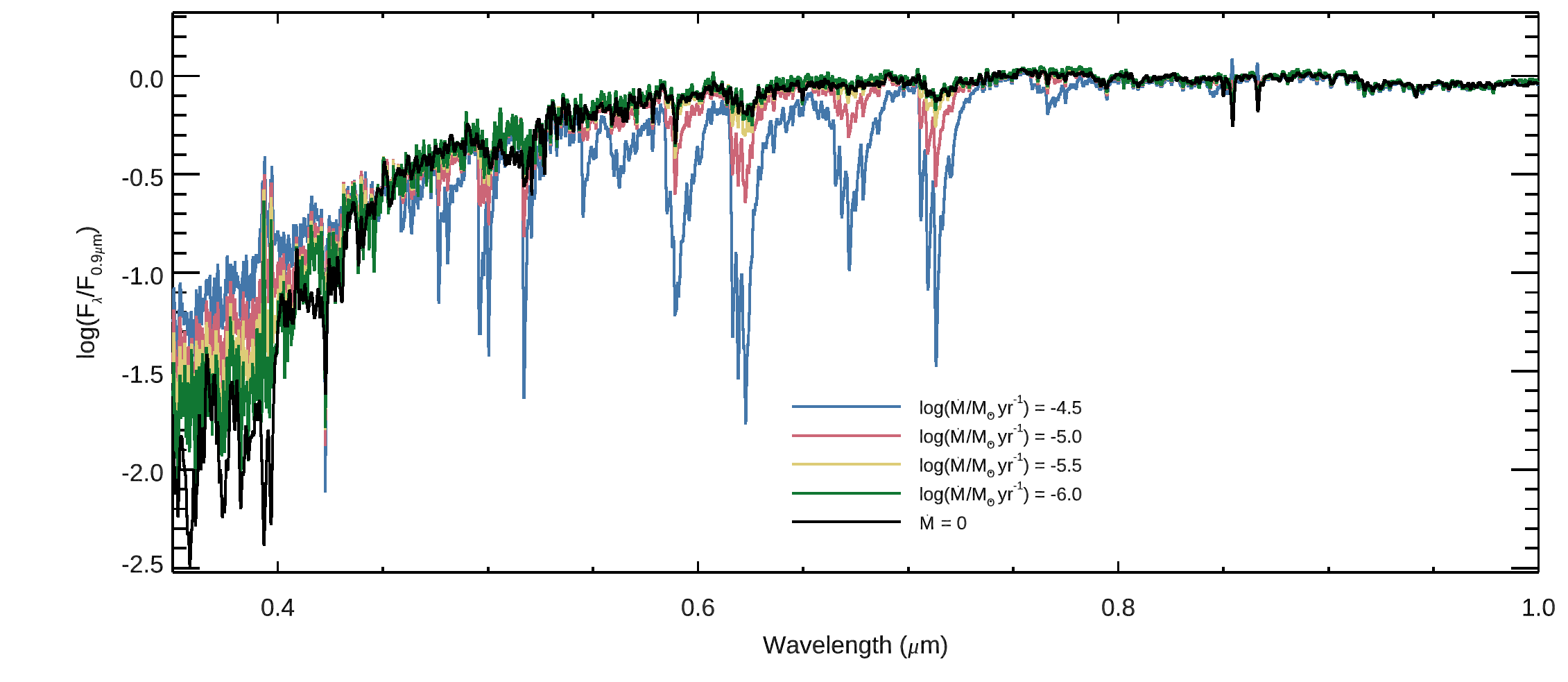}
\label{fig:speczoom}
\caption{As above, but a zoom-in of the near-UV/optical region.}
\end{center}
\end{figure*}

\begin{figure*}
\begin{center}
\includegraphics[width=17cm]{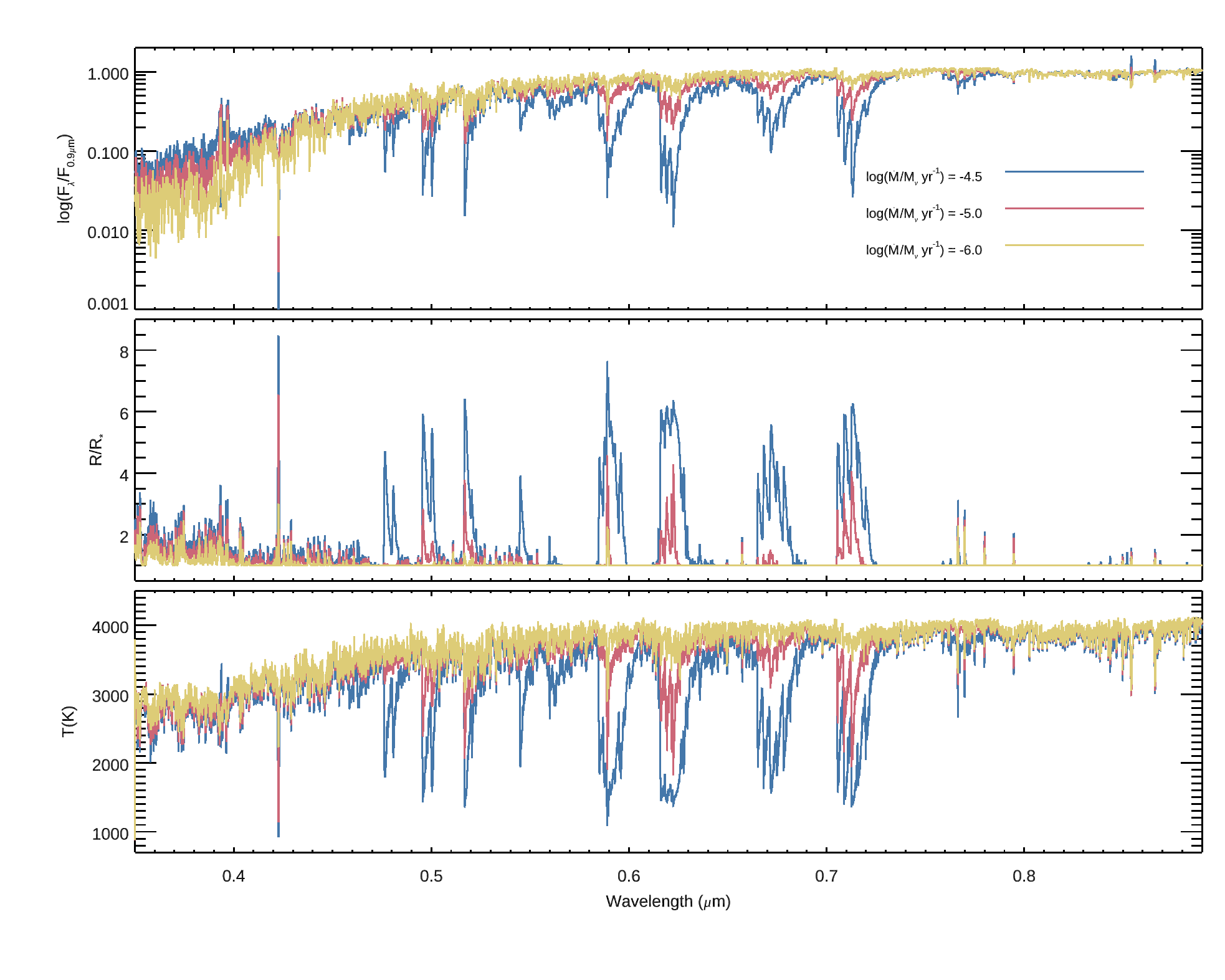}
\caption{The flux (top), radius of last-scattering surface (middle) and temperature at last-scattering surface (bottom) as a function of wavelength in the optical region of the spectrum for three mass-loss rates. As in \fig{fig:spectrum}, all models have \teff=3800\,K, \logg=0.0, Solar metallicity and Solar-scaled relative abundances. All spectra degraded to a spectral resolution of $R=2000$ for clarity.}
\label{fig:TR}
\end{center}
\end{figure*}

\subsection{Assumptions in the computation of spectra}
To compute the appearance of our models, we make a number of simplifying assumptions. 

\begin{itemize}
\item We assume the wind is in local thermodynamic equilibrium (LTE) at all depths. This then allows the straight-forward calculation of the occupation numbers of all atomic and molecular species, and hence the computation of the opacity as a function of depth and wavelength. A comprehensive discussion of the validity of the assumption of LTE in cool star atmospheres is presented in \citet{plez08}, but briefly the density in the outer layers is low enough for it to be unlikely that the gas is thermalized. However, at the current time collisional cross-sections for molecules are not yet known to a sufficient level to permit the calculation of fully non-LTE models of cool stars. Non-LTE corrections for atomic lines in near-IR spectra of RSG are known to be important when performing abundance analysis \citep[][]{bergemann12,bergemann13,bergemann15}, but result in only small changes ($\la$30\%) to equivalent widths, and are not considered here as we are mainly considering broad-band effects. In our investigation of how a wind affects the $J$-band spectra of RSGs (see Sect.\ \ref{sec:jband}), our work should be treated as comparative, i.e. fully LTE, with/without a wind. 
\item To simplify the radiative transfer we assume no velocity gradient -- that is, the velocity $v$ is zero at all depths. Our justification for this is that the terminal wind speed of RSGs is typically below the macroturbulent velocity \citep[$\sim 25$\kms, see ][]{Cunha07}. Hence, resonance zones for photon interactions extend almost the full depth of the wind. Since our goal in this work is to study the effect of a wind on the broad-band spectral appearance, this assumption is justifiable. The study of detailed line profiles would require the relaxing of this assumption.
\item We assume the same $T(r)$ for all mass-loss rates. To our knowledge, there is currently no first principles model that would allow us to estimate how Betelgeuse's chromospheric temperature profile would respond to increased circumstellar densities. In principle it may be possible to empirically calibrate this using a large sample of stars with known mass-loss rates. However, this is beyond the scope of this current work; instead here we focus on studying the qualitative variations of absorption line strength with mass-loss rate. 
\end{itemize}



To compute the emerging flux of our models, we adopt two methods. Firstly, to calculate the broad-band spectral energy distributions (SEDs) we use a modified version of the MARCS code. For detailed hi-resolution spectra, we use the code {\sc Turbospectrum} v19.1 \citep{plez12}. Both codes solve the radiative transfer equation in spherical symmetry. The {\sc Turbospectrum} code has previously been used to study the wind emission of cool stars \citep{Zamora14}. Following \citet{MCpaper}, we adopt a uniform microturbulent velocity of 4\kms, and a spectral sampling of 1.5\kms. In order to ensure that the outer wind is well-sampled in the spherical radiative transfer, we set the parameter {\tt diflog=1.0001}, which determines the ray spacing per unit impact parameter.

In both the MARCS and {\sc Turbospectrum} codes, we include the full suite of atomic and molecular data. Chemical equilibrium is solved for 92 atoms and their first 2 ions, and 519 molecules, including Ti, Ti$^+$, TiO and TiO$_2$. Condensation onto dust grains is not included, but is not expected to happen above 1000K for TiO at the pressures of our model winds of $< 10^-2 {\rm dyn\,cm^{-2}}$ \citep[cf.\ ][]{Woitke18}, which is consistent with boundary temperature for all of our models with mass-loss rates of \logmdot=-4 and lower.


\section{Results} \label{sec:results}

\subsection{Impact of winds on the spectrum}
In Figures \ref{fig:spectrum} and \ref{fig:speczoom} we plot the spectra of our model at various mass-loss rates, as well as the $\dot{M}=0$ (zero-wind) MARCS model. The observed behaviour is somewhat predictable; specifically, at higher mass-loss rates the density at a given local temperature in the inner wind is higher. This causes the opacity in the molecular lines (in particular the TiO lines between 0.5-0.8\um) to be higher, leading to enhanced absorption as a function of increasing \mdot. Correspondingly, we also see increased molecular emission in the near- and mid-IR, from CO and SiO, at higher mass-loss rates. The extra absorption in the optical would result in the star being classified with a much later spectral type, despite the fact that the underlying effective temperature remains unchanged. Specifically, the TiO band-strengths of the zero-wind model correspond to a spectral type of M0, while the models with \logmdot\ of -6, -5.5 and -5 would be classified as M1, M2 and >M5 respectively. 

In the near-UV/blue region we see an opposite effect: denser winds lead to brighter flux shortward of 0.45\um. This is {\it not} a continuum effect. Rather, it is the optical depth of the $G$-band and lines of several atomic species which increase, such that the last-scattering surface moves out into the chromospheric region where the temperature is higher than would be expected from radiative equilibrium. 

At modest \mdot\ ($\la 10^{-6}$\msunyr) we find that the above-described effect is small. There is slightly more absorption in the optical TiO bands and in the near-IR CO bands, whilst there is weak emission from CO at $\sim$4\um\ and SiO at $\sim$8\um. These effects gradually become more pronounced as \mdot\ is increased. By the time we reach \logmdot =-5, the TiO absorption has decreased the flux in the $V$ and $R$ bands by factors of several, and SiO/CO emission increases the flux in the mid-IR by a similar factor. We emphasise that this is {\it not} reddening by dust; we have not included dust in our model. The shift of flux away from the optical to the near-UV and IR is purely due to the various molecular bands in different spectral regions becoming more opaque at higher mass-loss rates.


The effect of mass-loss on the optical photosphere is explored in the middle panel of \fig{fig:TR}. At increasing \mdot, the surface of last-scattering becomes more inflated, and highly wavelength sensitive. At our highest \mdot, the TiO `photosphere' can be several times larger than that of the continuum photosphere, with local temperatures below 2000K. For high \mdot\ stars (\logmdot $\ga$ -5) optical angular diameter measurements may therefore overestimate the size of RSGs by a factor of a few (see also Sect.\ \ref{sec:MOLsphere}).

\begin{figure*}
\begin{center}
\includegraphics[width=17cm]{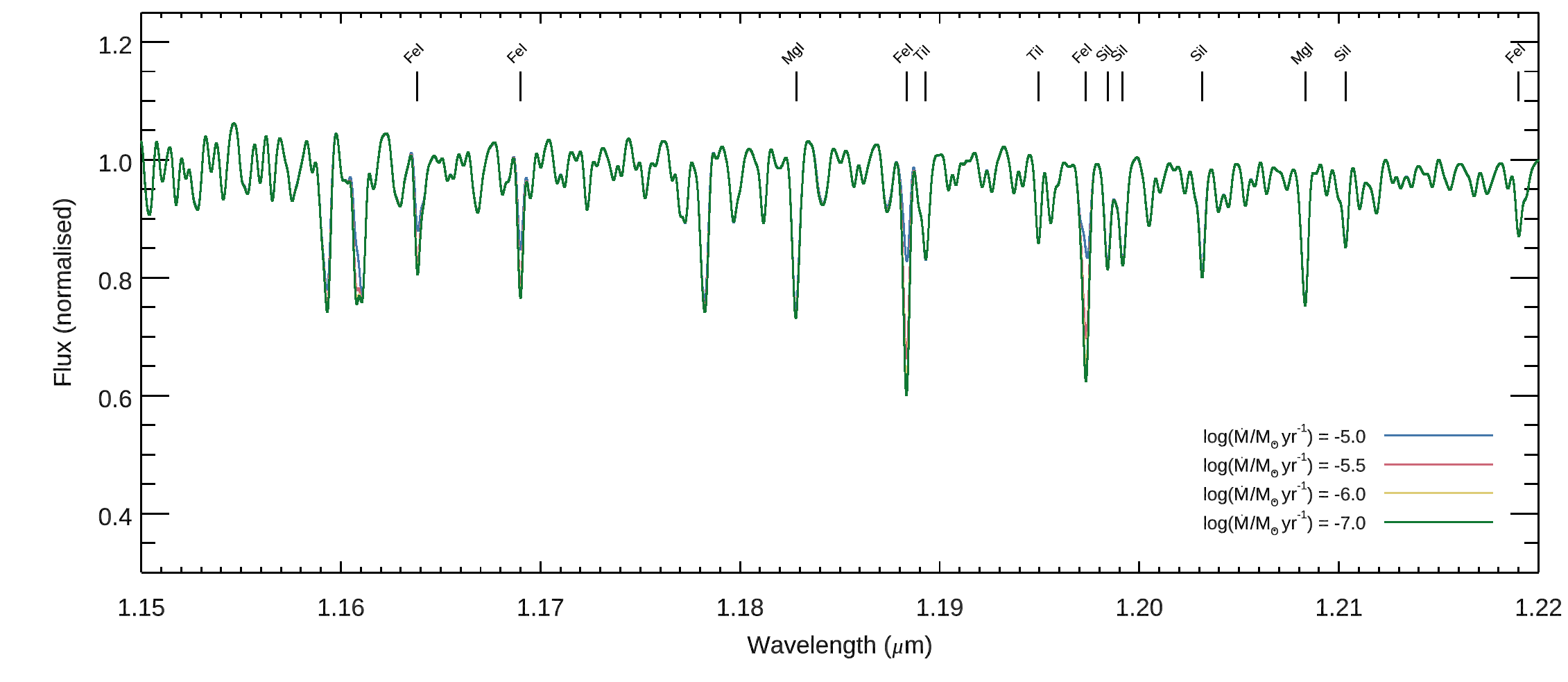}
\caption{The impact of winds on the $J$-band spectra of RSGs. The locations of diagnostic lines frequently used in the determination of stellar properties are indicated at the top of the plotting panel. All spectra have been degraded to a resolving power of $\lambda/\Delta\lambda = 4000$.}
\label{fig:jband}
\end{center}
\end{figure*}

\subsubsection{Optical vs.\ near-IR} \label{sec:jband}
Whilst the impact of winds on the optical spectrum is prominent, the effect in the near-IR bands is less so. Looking at \fig{fig:spectrum}, with the exception of a slight change to the CO features in the $H$ and $K$ bands, the broad-band fluxes appear largely unaltered. In \fig{fig:jband} we look in more detail at the portion of the $J$-band spectrum which has frequently been used to determine effective temperatures and metallicities \citep[e.g.][]{rsg_jband,MCpaper}. We find that, at \mdot\ $\la 10^{-6}$\msunyr, the $J$-band spectrum is almost completely unaltered. This is demonstrated by the fact that the green and yellow lines in \fig{fig:jband} are indistinguishable at almost all wavelengths. At the higher mass-loss rate, the strengths of the diagnostic lines are affected only slightly, with the exception of the {Fe {\sc i}} lines which become weaker at very high mass-loss rates as they start to be filled in by emission in the cores of the lines. These results suggest that RSG fundamental parameters derived from $J$-band spectra should be robust to whether or not a wind is included in the analysis, though systematic effects may come into play at spectral types later than $\sim$M2. 

\begin{figure*}
\begin{center}
\includegraphics[width=17cm]{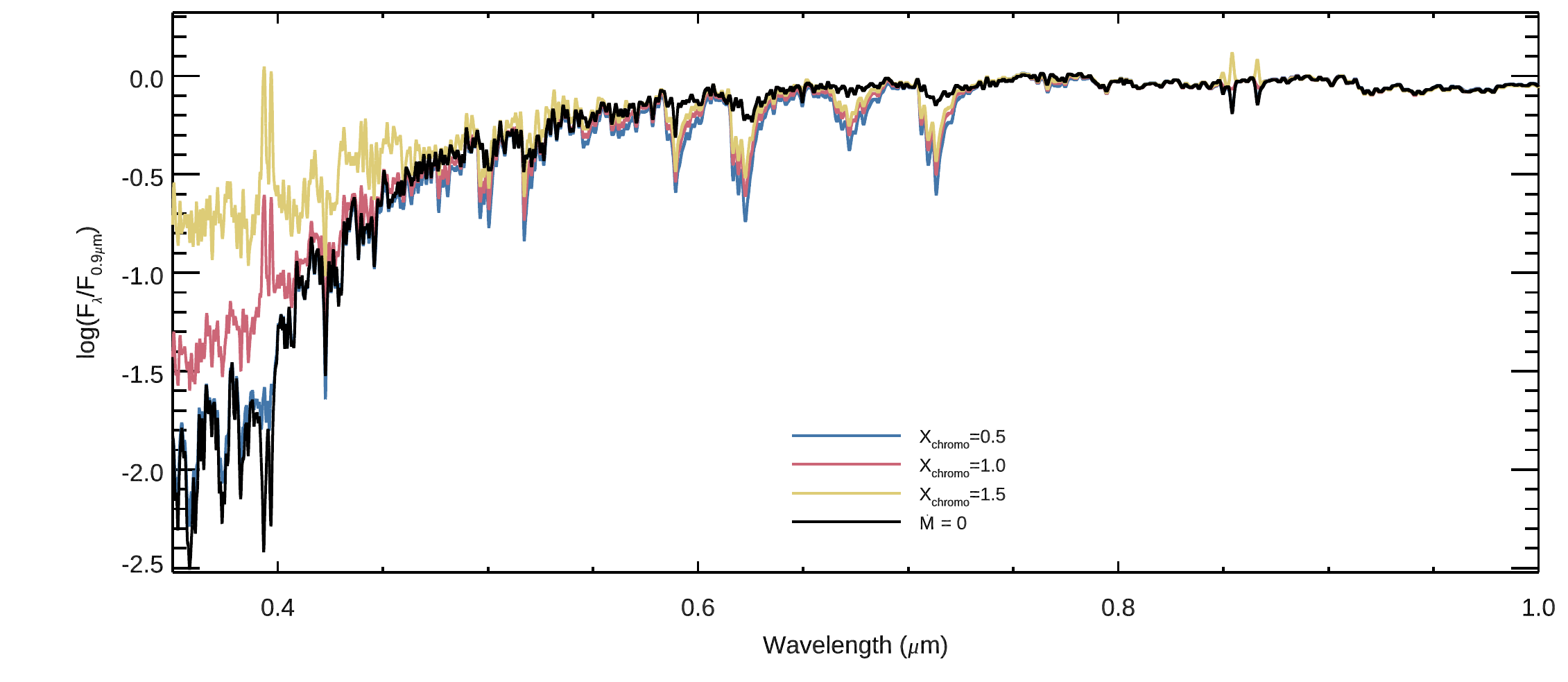}
\caption{The effect of modifying the peak chromospheric temperature of the \logmdot=-5 model on its broadband optical spectrum. The spectral resolution has been degraded to $R=500$ for clarity. }
\label{fig:chromo}
\end{center}
\end{figure*}

\subsubsection{Effect of varying the chromospheric temperature} \label{sec:chromo}
In all models discussed thus far, the temperature profile in the chromosphere is motivated by the semi-empirical model of Betelgeuse presented in \citet{Harper01}. As already mentioned, we do not know how the chromosphere would respond to changes in mass-loss rate, so in this section we perform the exploratory exercise of varying the scale of the temperature inversion to see how the resulting spectrum responds. We do this by multiplying the temperature peak by a factor  $X_{\rm chromo}$, whilst ensuring that the temperature either side of the inversion (i.e. at 1.1 < $R/R_\star < 5.0$) remains constant. We do this for two somewhat extreme values of $X_{\rm chromo} = \{0.5,1.5\}$, which lead to peak chromospheric temperatures of 2000K and 6000K respectively. The fiducial model (shown as black triangles in Fig.\ \ref{fig:modprof}) is described by a value of $X_{\rm chromo} = 1$; while the $X_{\rm chromo} = 0.5$ model approximates a temperature profile consistent with radiative equilibrium (shown as blue crosses in Fig. \ref{fig:modprof}). 

In Fig.\ \ref{fig:chromo} we plot the optical spectra of the above described tests for a mass-loss rate of \logmdot=-5.0, along with the zero-wind model for comparison. We find that modifying the chromospheric temperature can alter the absorption balance of the different atomic and molecular species. In the coolest chromospheric model ($X_{\rm chromo} = 0.5$, effectively the `radiative' temperature profile of Fig.\ 1, plotted in blue in Fig.\ \ref{fig:chromo}) we see slightly enhanced TiO absorption with respect to the fiducial model (plotted in red). This is caused by the lower temperatures throughout the wind leading to an increased abundance of TiO close to the star. However, the flux in the blue region of the spectrum, which forms close to the chromosphere, is reduced to lower values of $X_{\rm chromo}$ due to the $T^4$ dependence. In the hotter chromosphere model ($X_{\rm chromo} = 1.5$, which corresponds to a peak chromosphere temperature of 6000K, plotted in yellow in Fig.\ \ref{fig:chromo}), we see the reverse -- slightly weaker TiO absorption, but increased flux in the blue. In this particular model, we also see the calcium lines at $\sim$0.4\um\ and $\sim$0.87\um\ go into emission. 

From the above-described experiments, we conclude that modifying the chromospheric temperature of our atmosphere+wind models has the effect of altering the flux in the blue whilst keeping the rest of the optical spectrum approximately steady. This is important for two reasons. Firstly, fluctuations in the chromospheric temperature could cause variations in the star's $B-V$ colour, without altering the star's effective temperature. Secondly, the inclusion of a chromosphere in these `windy' models naturally explains the `blue-excess' reported in RSG spectra when compared to a MARCS model tuned to fit the TiO bands \citep[e.g.][]{massey05}. We will return to these issues in Sect.\ \ref{sec:betel}.

\begin{figure*}
\begin{center}
\includegraphics[width=17cm]{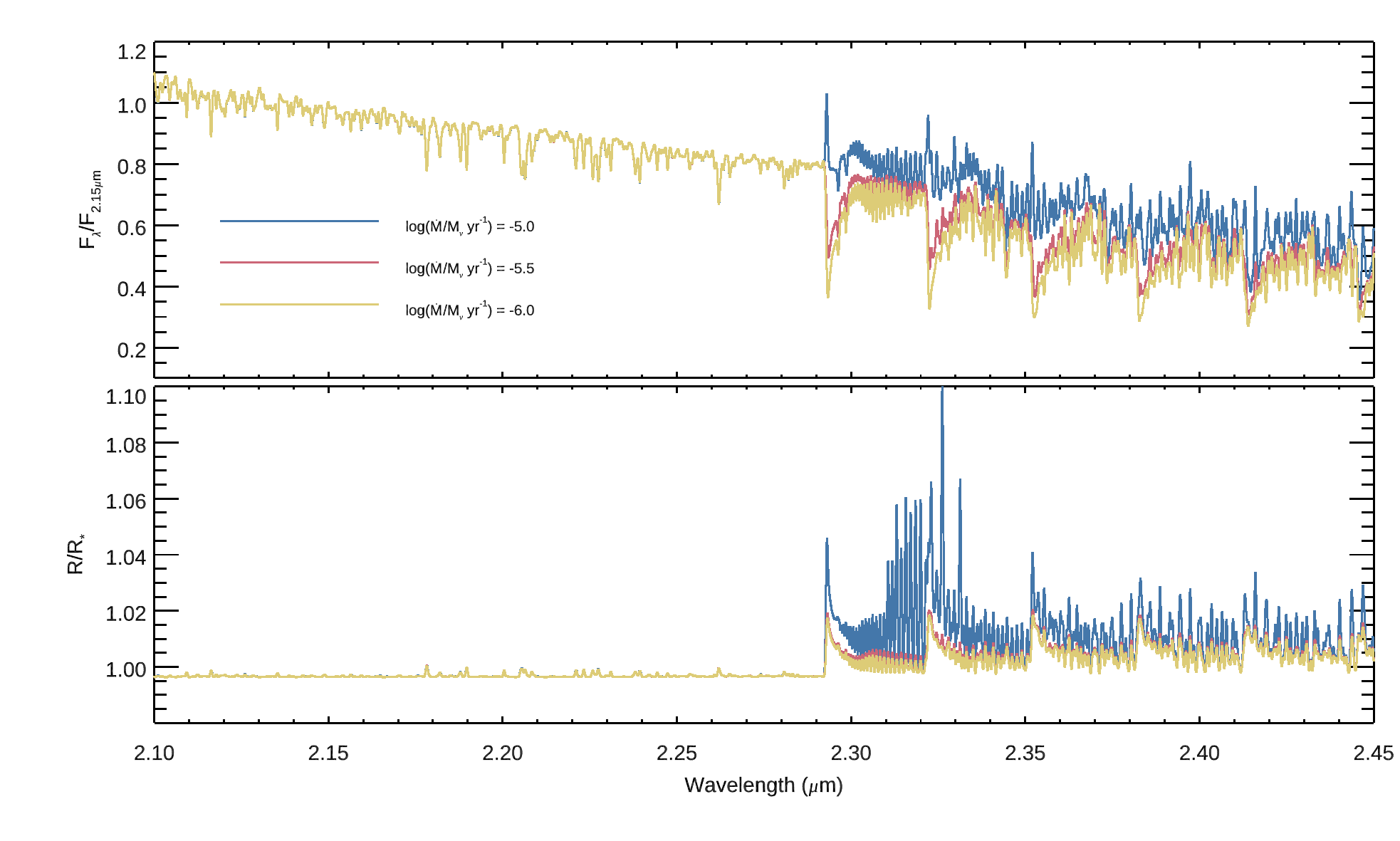}
\caption{The spectra of our models in the region of the 2.3\um\ CO bandhead (top), and the radius of the $\tau=1$ surface as a function of wavelength (bottom). Spectra have been degraded to a resolving power of $\lambda/\Delta\lambda = 4000$. The predicted signature of a spatially extended CO forming region is similar to that observed in $\alpha$~Ori, which has been previously explained by the presence of a MOLsphere. }
\label{fig:CO}
\end{center}
\end{figure*}

\section{Discussion} \label{sec:disc}
As shown in the previous Section, the inclusion of a wind in RSG model atmospheres has the potential to substantially alter the spectral morphology of the star. In this Section, we discuss the implications of these results in the context of several outstanding problems and questions pertaining to RSGs.

\subsection{Unification of optical and near-IR temperatures} \label{sec:optir}
The most widely used temperature scales for RSGs are those presented in \citet{Levesque05,Levesque06}. These authors estimated temperatures as a function of spectral type by fitting model spectra obtained from MARCS (zero-wind) atmospheres to the TiO bands in the optical regions of RSGs. Despite the apparent improved agreement between their updated temperature scale and the temperatures of RSGs in evolutionary calculations, there were several inconsistencies in their results. Firstly, in many of their spectral fits, it can be seen that the model which fits the TiO bands fails to reproduce any other part of the spectrum. For example, their fit to $\alpha$~Her \citep[see ][Fig.\ 1.1]{Levesque05} matches the TiO bands perfectly, but underestimates the flux below 0.5\um\ and above 0.8\um\ by a factor of 2-3. Secondly, temperatures estimated from $V-K$ colours are systematically higher than those of the model fits to the optical by several hundred Kelvin (Fig.\ 5a in \cite{Levesque05} and Fig.\ 4 in \cite{Levesque06}).
Thirdly, the authors found a population of stars whose temperatures (derived from the TiO band strengths) seemed to violate the Hayashi limit \citep{Levesque07}.

In \citet{rsgteff}, the methodology of estimating \teff\ from the TiO bands was challenged. Specifically, it was shown that estimating the \teff\ from fits to the near-IR continuum resulted in systematically higher temperatures than those inferred from the TiO bands. Furthermore, it was shown that tuning MARCS (zero-wind) model atmospheres to the TiO bands resulted in a spectral fit that failed to reproduce any other spectral feature. Conversely, the fit to the near-IR continuum performed much better at all wavelengths, but produced a poor fit to the TiO bands. Whilst Davies et al.\ presented arguments that their warmer RSG temperatures were more reliable, it is clear from both that work and the Levesque et al.\ papers that the MARCS model atmospheres are missing a key physical ingredient, and as such the temperatures of RSGs estimated from these models should be viewed with skepticism. 

The results of our study presented here have shown that by adding a simple wind to a static MARCS model atmosphere, with a mass-loss rate which is comparable to all contemporary observations, it is possible to increase the strengths of the TiO bands whilst keeping the near-IR continuum fixed. Furthermore, adding a wind can also decrease the flux in the $V$ and $R$ bands whilst keeping it steady in the blue (see top panel of Fig.\ \ref{fig:TR}), explaining the 'blue-excess' seen in many of the Levesque et al.\ stars. Hence, by adding winds to RSG model atmospheres, it is now possible to fit all spectral features from the near-UV to the IR simultaneously.

\subsection{Correlation of spectral type with IR excess} \label{sec:spt-mdot}
Several authors have noted the apparent correlation between spectral type and excess flux in the mid-IR and/or some other marker of dense circumstellar material (e.g. such as OH maser emission which seems to be confined to RSGs with late spectral types) \citep{vanLoon05,Negueruela13,Gonzalez-Fernandez15,Dorda16,Beasor-Davies16,Beasor-Davies18,Davies-Beasor18}. Several of these authors speculated that late spectral types in RSGs are indicative of advanced evolutionary state, when presumably the mass-loss rate is greater and/or the star has had more time to accumulate circumstellar material. The results of this work suggest that the link between spectral type and circumstellar material is more direct; that the increased absorption in the TiO bands, the increased circumstellar density, and hence the increased IR excess are all signatures of a high mass-loss rate.


\subsection{The dimming of Betelgeuse} \label{sec:betel}
During 2018-2020, the nearby RSG Betelgeuse famously underwent a dimming in the optical of around a magnitude \citep{Dupree20}. During this time, the flux in the near-IR changed only very little \citep{Gehrz20,Guinan-Wasatonic20}, while the spectral type became later as the V-band flux dropped  \citep{Harper20}. The star was substantially variable in the UV during this time \citep{Dupree20}, while a high resolution $V$-band image of Betelgeuse from VLT+SPHERE shows that an optically-dark `patch'  appeared over a patch of the star during the dimming \citep{Montarges21}. 

In terms of an explanation for the dimming event, a varying effective temperature scenario is ruled out by time-resolved angular size measurements; while a transiting dust clump scenario can be eliminated by high-resolution time-series imaging of the dark spot \citep{Montarges21}. The remaining possible explanations fall into roughly two categories; an increased opacity in the outer atmosphere from either (a) newly-formed dust or (b) molecules. Evidence for the latter comes from time-resolved TiO band monitoring and tomography \citep{Harper20,Kravchenko21}. High-resolution polarimetry data do not seem to conclusively distinguish between the two \citep{Cotton20,Safonov20}, nor does high resolution imaging \citep{Montarges21}. Ultraviolet observations indicate the formation of a dense structure in the chromosphere just before the onset of the dimming, indicating possibly the formation of the feature that would go on to obscure the star \citep{Dupree20}, but cannot determine what the obscuring feature is. 

The first claim for the `dust' scenario was presented in \citet{Levesque-Massey20}, who compared optical spectrophotometry from 2020 with similar data taken in 2004, long before the dimming began. In the 2020 spectrum, they found excess flux in the blue with respect to their best-fitting MARCS model, and interpreted this as being caused by grey dust scattering from large dust grains. However, the explanation for the `blue-excess' in the 2020 spectrum could equally be explained by incorrect opacity in their (zero-wind) MARCS model atmosphere. We have shown in Sect.\ \ref{sec:chromo} and in Fig.\ \ref{fig:chromo} that the flux in the blue is sensitive both to mass-loss and the temperature at the base of the wind; and indeed the effect of a temperature inversion in the inner wind mimics the `blue-excess' phenomenon almost perfectly. Furthermore, the explanation of dust-scattering for the blue excess, which must be strongly wavelength dependent if it appears only at wavelengths $<0.4$\um, is inconsistent with the further conclusion in \citet{Levesque-Massey20} that the dust extinction must also be grey.

From the results of our work, we offer an alternative explanation for the dimming of Betelgeuse. The ejected material observed in the UV by \citet{Dupree20} must gradually cool as it moves away from the star. However, before this gas blob can condense into dust, it must first form molecules. As the blob cools to below $\sim$3000K, the optical depth in the TiO bands will dramatically increase, causing a dimming in the optical region of the spectrum whilst leaving the near-IR largely unaffected \citep[see also ][]{Harper20,Kravchenko21}.

In principle, one could tailor the density and temperature profiles of our model to tune the absorption strengths of individual molecular species (see Sect.\ \ref{sec:chromo}), and provide tailored quantitative fits to Betelgeuse's light-curve and colour evolution throughout the dimming. However, we choose not to do that since ultimately we would be using a 1-D model to fit a phenomenon that we know is inherently 3-D in nature. Instead, we simply conclude that molecular opacity fluctuations caused by the ejection of a dense blob of gas are capable of explaining the colour and brightness changes of Betelgeuse without changing the star's effective temperature or circumstellar dust mass.

\subsection{`MOLspheres'} \label{sec:MOLsphere}
It has been recognised since the 1970s that there are several spectroscopic features in the near- and mid-IR that cannot be reproduced by a static atmosphere \citep[for a review, see][]{Tsuji06}. More recently, interferometry has shown that RSGs have wavelength-dependent photospheres, appearing to be larger at wavelengths co-incident with high molecular opacities \citep[e.g.][]{Perrin04,Perrin05,Montarges14}. These phenomena are frequently explained by the ad-hoc addition of a static molecular shell (or `MOLsphere') above the hydrostatic photosphere \citep{tsuji00,Tsuji06}, though it is not known what gives rise to the existence of such a structure. 

The addition of a wind to a static model atmosphere could reproduce many of the observed features of a MOLsphere. The temperature and density profiles of the \citet{Harper01} model, scaled to a typical RSG mass-loss rate, naturally produce the appropriate circumstellar densities and temperatures. Moreover, the very slow acceleration of the wind means that the inner $\sim$1\rstar\ of material is almost static with respect to macroturbulent motions. In Fig.\ \ref{fig:CO}, we give an example of how the measured angular size of a RSG in the $K$-band might vary as a function of wavelength for different mass-loss rates. At higher \mdot, the opacity of the 2.3\um\ CO band increases, causing the star to appear increasingly larger at those wavelengths. The qualitative agreement of this figure with Fig.\ 5 of \citet{Montarges14} is extremely encouraging. At this stage we do not attempt to provide a quantitative match to observations, but it is possible that by fine-tuning e.g. the density and temperature structure of our model we would be able to reproduce many of the spectral features associated with MOLspheres.

\section{Conclusions} \label{sec:conc}

We have presented the first synthesis of the entire optical/near-IR spectrum of Red Supergiants using model atmospheres that include a wind. Our main findings are summarised as follows:

\begin{itemize}
\item The addition of a wind increases the amount of TiO absorption at fixed effective temperature \teff. For mass-loss rates \mdot\ at the higher end of that observed for RSGs ($\sim 10^{-5}$\msunyr) the effect is dramatic, shifting the star's spectral type from M0 to M5, despite \teff\ remaining constant.
\item Though the flux in the {\it VRI} bands reduces with increased \mdot, the flux in the $U$ and $B$ bands can {\it increase}. The magnitude of this blue-excess is sensitive to the temperature in the chromosphere as well as the mass-loss rate. 
\item High mass-loss rates also produce mid-IR excess from the molecular bands of SiO and CO. This excess is independent of any dust emission.
\end{itemize}

Our results have several implications for the understanding of RSGs as they go some way towards explaining various observational phenomena:

\begin{itemize}
\item It is well known that zero-wind model atmospheres cannot simultaneously match the optical and near-IR spectrum, with cooler effective temperatures required in the optical to match the TiO bands. The inclusion of a wind naturally explains this, as it increases the TiO absorption without altering the near-IR.
\item Our atmosphere+wind models also provide an explanation for the `blue-excess' seen in some RSGs, whereby a star is observed to be brighter in the blue compared to a (zero-wind) MARCS model tuned to fit the TiO bands. Rather than being caused by dust scattering, the blue-excess results from increased circumstellar densities coupled with a chromosphere-type temperature inversion.
\item It has also been noted that RSGs with dense circumstellar material, as evidenced by large amounts of mid-IR excess and/or maser emission, are usually also found to have late spectral types. Furthermore, RSGs in star clusters appear to show a clear correlation between mid-IR excess and spectral type. Again, this is naturally explained by a wind: high mass-loss rates result in both a dense CSM {and} strong TiO absorption.
\item The recent dimming of Betelgeuse, in which the optical flux dipped by an order of magnitude while the near-IR flux stayed almost constant, can be explained as increased TiO absorption at roughly constant effective temperature. This is what is expected from a period of increased mass-loss, which is consistent with UV spectroscopy and high spatial resolution imaging of Betelgeuse. 
\item Finally, the addition of a wind, when adopting the semi-empirical model for that of Betelgeuse, reproduces many of the observed features of RSGs that have typically been attributed to `MOLspheres'. Therefore, a slowly accelerating wind may be the explanation for the extended molecular zone which has been inferred to exist around many RSGs. 
\end{itemize}

\section*{Data availability}
The data underlying this article are available at the Centre de Données astronomiques de Strasbourg (CDS), and can be accessed under the catalogue identifier J/MNRAS/<vol>/<page>.

\section*{Acknowledgements} 
We are very grateful to Miguel Montarges for discussions relating to the dimming of Betelgeuse. We also thank Nathan Smith and Emma Beasor for useful discussions. Finally, we thank the anonymous referee for their suggestions and comments that helped us improve our manuscript. This work made use of the IDL astronomy library, available at {\tt https://idlastro.gsfc.nasa.gov}, and the Coyote IDL graphics library. 




\bibliographystyle{mnras}
\bibliography{biblio} 

\begin{thebibliography}{}
\makeatletter
\relax
\def\mn@urlcharsother{\let\do\@makeother \do\$\do\&\do\#\do\^\do\_\do\%\do\~}
\def\mn@doi{\begingroup\mn@urlcharsother \@ifnextchar [ {\mn@doi@}
  {\mn@doi@[]}}
\def\mn@doi@[#1]#2{\def\@tempa{#1}\ifx\@tempa\@empty \href
  {http://dx.doi.org/#2} {doi:#2}\else \href {http://dx.doi.org/#2} {#1}\fi
  \endgroup}
\def\mn@eprint#1#2{\mn@eprint@#1:#2::\@nil}
\def\mn@eprint@arXiv#1{\href {http://arxiv.org/abs/#1} {{\tt arXiv:#1}}}
\def\mn@eprint@dblp#1{\href {http://dblp.uni-trier.de/rec/bibtex/#1.xml}
  {dblp:#1}}
\def\mn@eprint@#1:#2:#3:#4\@nil{\def\@tempa {#1}\def\@tempb {#2}\def\@tempc
  {#3}\ifx \@tempc \@empty \let \@tempc \@tempb \let \@tempb \@tempa \fi \ifx
  \@tempb \@empty \def\@tempb {arXiv}\fi \@ifundefined
  {mn@eprint@\@tempb}{\@tempb:\@tempc}{\expandafter \expandafter \csname
  mn@eprint@\@tempb\endcsname \expandafter{\@tempc}}}

\bibitem[\protect\citeauthoryear{{Beasor} \& {Davies}}{{Beasor} \&
  {Davies}}{2016}]{Beasor-Davies16}
{Beasor} E.~R.,  {Davies} B.,  2016, \mn@doi [\mnras] {10.1093/mnras/stw2054},
  \href {http://adsabs.harvard.edu/abs/2016MNRAS.463.1269B} {463, 1269}

\bibitem[\protect\citeauthoryear{{Beasor} \& {Davies}}{{Beasor} \&
  {Davies}}{2018}]{Beasor-Davies18}
{Beasor} E.~R.,  {Davies} B.,  2018, \mn@doi [\mnras] {10.1093/mnras/stx3174},
  \href {http://adsabs.harvard.edu/abs/2018MNRAS.475...55B} {475, 55}

\bibitem[\protect\citeauthoryear{{Beasor}, {Davies}, {Smith}, {van Loon},
  {Gehrz}  \& {Figer}}{{Beasor} et~al.}{2020}]{Beasor20}
{Beasor} E.~R.,  {Davies} B.,  {Smith} N.,  {van Loon} J.~T.,  {Gehrz} R.~D.,
  {Figer} D.~F.,  2020, \mn@doi [\mnras] {10.1093/mnras/staa255}, \href
  {https://ui.adsabs.harvard.edu/abs/2020MNRAS.492.5994B} {492, 5994}

\bibitem[\protect\citeauthoryear{{Bergemann}, {Kudritzki}, {Plez}, {Davies},
  {Lind}  \& {Gazak}}{{Bergemann} et~al.}{2012}]{bergemann12}
{Bergemann} M.,  {Kudritzki} R.-P.,  {Plez} B.,  {Davies} B.,  {Lind} K.,
  {Gazak} Z.,  2012, \mn@doi [\apj] {10.1088/0004-637X/751/2/156}, \href
  {http://adsabs.harvard.edu/abs/2012ApJ...751..156B} {751, 156}

\bibitem[\protect\citeauthoryear{{Bergemann}, {Kudritzki}, {W{\"u}rl}, {Plez},
  {Davies}  \& {Gazak}}{{Bergemann} et~al.}{2013}]{bergemann13}
{Bergemann} M.,  {Kudritzki} R.-P.,  {W{\"u}rl} M.,  {Plez} B.,  {Davies} B.,
  {Gazak} Z.,  2013, \mn@doi [\apj] {10.1088/0004-637X/764/2/115}, \href
  {http://adsabs.harvard.edu/abs/2013ApJ...764..115B} {764, 115}

\bibitem[\protect\citeauthoryear{{Bergemann}, {Kudritzki}, {Gazak}, {Davies}
  \& {Plez}}{{Bergemann} et~al.}{2015}]{bergemann15}
{Bergemann} M.,  {Kudritzki} R.-P.,  {Gazak} Z.,  {Davies} B.,   {Plez} B.,
  2015, \mn@doi [\apj] {10.1088/0004-637X/804/2/113}, \href
  {http://adsabs.harvard.edu/abs/2015ApJ...804..113B} {804, 113}

\bibitem[\protect\citeauthoryear{{Carpenter}, {Robinson}, {Harper}, {Bennett},
  {Brown}  \& {Mullan}}{{Carpenter} et~al.}{1999}]{Carpenter99}
{Carpenter} K.~G.,  {Robinson} R.~D.,  {Harper} G.~M.,  {Bennett} P.~D.,
  {Brown} A.,   {Mullan} D.~J.,  1999, \mn@doi [\apj] {10.1086/307520}, \href
  {https://ui.adsabs.harvard.edu/abs/1999ApJ...521..382C} {521, 382}

\bibitem[\protect\citeauthoryear{{Chiosi} \& {Maeder}}{{Chiosi} \&
  {Maeder}}{1986}]{Chiosi-Maeder86}
{Chiosi} C.,  {Maeder} A.,  1986, \mn@doi [\araa]
  {10.1146/annurev.aa.24.090186.001553}, \href
  {http://adsabs.harvard.edu/abs/1986ARA%26A..24..329C} {24, 329}

\bibitem[\protect\citeauthoryear{{Cotton}, {Bailey}, {De Horta}, {Norris}  \&
  {Lomax}}{{Cotton} et~al.}{2020}]{Cotton20}
{Cotton} D.~V.,  {Bailey} J.,  {De Horta} A.~Y.,  {Norris} B. R.~M.,   {Lomax}
  J.~R.,  2020, \mn@doi [Research Notes of the American Astronomical Society]
  {10.3847/2515-5172/ab7f2f}, \href
  {https://ui.adsabs.harvard.edu/abs/2020RNAAS...4...39C} {4, 39}

\bibitem[\protect\citeauthoryear{{Cunha}, {Sellgren}, {Smith}, {Ramirez},
  {Blum}  \& {Terndrup}}{{Cunha} et~al.}{2007}]{Cunha07}
{Cunha} K.,  {Sellgren} K.,  {Smith} V.~V.,  {Ramirez} S.~V.,  {Blum} R.~D.,
  {Terndrup} D.~M.,  2007, \mn@doi [\apj] {10.1086/521813}, \href
  {http://adsabs.harvard.edu/abs/2007ApJ...669.1011C} {669, 1011}

\bibitem[\protect\citeauthoryear{{Davies} \& {Beasor}}{{Davies} \&
  {Beasor}}{2018}]{Davies-Beasor18}
{Davies} B.,  {Beasor} E.~R.,  2018, \mn@doi [\mnras] {10.1093/mnras/stx2734},
  \href {http://adsabs.harvard.edu/abs/2018MNRAS.474.2116D} {474, 2116}

\bibitem[\protect\citeauthoryear{{Davies}, {Kudritzki}  \& {Figer}}{{Davies}
  et~al.}{2010}]{rsg_jband}
{Davies} B.,  {Kudritzki} R.,   {Figer} D.~F.,  2010, \mn@doi [\mnras]
  {10.1111/j.1365-2966.2010.16965.x}, \href
  {http://adsabs.harvard.edu/abs/2010MNRAS.407.1203D} {407, 1203}

\bibitem[\protect\citeauthoryear{{Davies} et~al.,}{{Davies}
  et~al.}{2013}]{rsgteff}
{Davies} B.,  et~al., 2013, \mn@doi [\apj] {10.1088/0004-637X/767/1/3}, \href
  {http://adsabs.harvard.edu/abs/2013ApJ...767....3D} {767, 3}

\bibitem[\protect\citeauthoryear{{Davies}, {Kudritzki}, {Gazak}, {Plez},
  {Bergemann}, {Evans}  \& {Patrick}}{{Davies} et~al.}{2015}]{MCpaper}
{Davies} B.,  {Kudritzki} R.-P.,  {Gazak} Z.,  {Plez} B.,  {Bergemann} M.,
  {Evans} C.,   {Patrick} L.,  2015, \mn@doi [ApJ]
  {10.1088/0004-637X/806/1/21}, \href
  {http://adsabs.harvard.edu/abs/2015ApJ...806...21D} {806, 21}

\bibitem[\protect\citeauthoryear{{De Beck}, {Decin}, {de Koter}, {Justtanont},
  {Verhoelst}, {Kemper}  \& {Menten}}{{De Beck} et~al.}{2010}]{debeck10}
{De Beck} E.,  {Decin} L.,  {de Koter} A.,  {Justtanont} K.,  {Verhoelst} T.,
  {Kemper} F.,   {Menten} K.~M.,  2010, \mn@doi [\aap]
  {10.1051/0004-6361/200913771}, \href
  {http://adsabs.harvard.edu/abs/2010A%26A...523A..18D} {523, A18}

\bibitem[\protect\citeauthoryear{{Decin}, {Hony}, {de Koter}, {Justtanont},
  {Tielens}  \& {Waters}}{{Decin} et~al.}{2006}]{decin06}
{Decin} L.,  {Hony} S.,  {de Koter} A.,  {Justtanont} K.,  {Tielens}
  A.~G.~G.~M.,   {Waters} L.~B.~F.~M.,  2006, \mn@doi [\aap]
  {10.1051/0004-6361:20065230}, \href
  {http://adsabs.harvard.edu/abs/2006A%26A...456..549D} {456, 549}

\bibitem[\protect\citeauthoryear{{Dorda}, {Negueruela},
  {Gonz{\'a}lez-Fern{\'a}ndez}  \& {Tabernero}}{{Dorda} et~al.}{2016}]{Dorda16}
{Dorda} R.,  {Negueruela} I.,  {Gonz{\'a}lez-Fern{\'a}ndez} C.,   {Tabernero}
  H.~M.,  2016, \mn@doi [\aap] {10.1051/0004-6361/201528024}, \href
  {https://ui.adsabs.harvard.edu/abs/2016A&A...592A..16D} {592, A16}

\bibitem[\protect\citeauthoryear{{Dupree} et~al.,}{{Dupree}
  et~al.}{2020}]{Dupree20}
{Dupree} A.~K.,  et~al., 2020, \mn@doi [\apj] {10.3847/1538-4357/aba516}, \href
  {https://ui.adsabs.harvard.edu/abs/2020ApJ...899...68D} {899, 68}

\bibitem[\protect\citeauthoryear{{Gehrz}, {Marchetti}, {McMillan}, {Procter},
  {Zarling}, {Bartlett}  \& {Smith}}{{Gehrz} et~al.}{2020}]{Gehrz20}
{Gehrz} R.~D.,  {Marchetti} J.,  {McMillan} S.,  {Procter} T.,  {Zarling} A.,
  {Bartlett} J.,   {Smith} N.,  2020, The Astronomer's Telegram, \href
  {https://hes32-ctp.trendmicro.com:443/wis/clicktime/v1/query?url=https%3a%2f%2fui.adsabs.harvard.edu%2fabs%2f2020ATel13518....1G&umid=c35e2bcf-7f3f-40cc-9a37-331445d9ef40&auth=768f192bba830b801fed4f40fb360f4d1374fa7c-e4a2571544e05d5cdd003d4d583f3cae0d6038f8}
  {13518, 1}

\bibitem[\protect\citeauthoryear{{Gonz{\'a}lez-Fern{\'a}ndez}, {Dorda},
  {Negueruela}  \& {Marco}}{{Gonz{\'a}lez-Fern{\'a}ndez}
  et~al.}{2015}]{Gonzalez-Fernandez15}
{Gonz{\'a}lez-Fern{\'a}ndez} C.,  {Dorda} R.,  {Negueruela} I.,   {Marco} A.,
  2015, \mn@doi [\aap] {10.1051/0004-6361/201425362}, \href
  {http://adsabs.harvard.edu/abs/2015A%26A...578A...3G} {578, A3}

\bibitem[\protect\citeauthoryear{{Groenewegen}, {Sloan}, {Soszy{\'n}ski}  \&
  {Petersen}}{{Groenewegen} et~al.}{2009}]{Groenewegen09}
{Groenewegen} M.~A.~T.,  {Sloan} G.~C.,  {Soszy{\'n}ski} I.,   {Petersen}
  E.~A.,  2009, \mn@doi [\aap] {10.1051/0004-6361/200912678}, \href
  {http://adsabs.harvard.edu/abs/2009A%26A...506.1277G} {506, 1277}

\bibitem[\protect\citeauthoryear{{Guinan} \& {Wasatonic}}{{Guinan} \&
  {Wasatonic}}{2020}]{Guinan-Wasatonic20}
{Guinan} E.~F.,  {Wasatonic} R.~J.,  2020, The Astronomer's Telegram, \href
  {https://ui.adsabs.harvard.edu/abs/2020ATel13439....1G} {13439, 1}

\bibitem[\protect\citeauthoryear{{Gustafsson}, {Edvardsson}, {Eriksson},
  {J{\o}rgensen}, {Nordlund}  \& {Plez}}{{Gustafsson}
  et~al.}{2008}]{gustafsson08}
{Gustafsson} B.,  {Edvardsson} B.,  {Eriksson} K.,  {J{\o}rgensen} U.~G.,
  {Nordlund} {\AA}.,   {Plez} B.,  2008, \mn@doi [\aap]
  {10.1051/0004-6361:200809724}, \href
  {http://adsabs.harvard.edu/abs/2008A%26A...486..951G} {486, 951}

\bibitem[\protect\citeauthoryear{{Harper}, {Brown}  \& {Lim}}{{Harper}
  et~al.}{2001}]{Harper01}
{Harper} G.~M.,  {Brown} A.,   {Lim} J.,  2001, \mn@doi [\apj]
  {10.1086/320215}, \href
  {https://ui.adsabs.harvard.edu/abs/2001ApJ...551.1073H} {551, 1073}

\bibitem[\protect\citeauthoryear{{Harper}, {Guinan}, {Wasatonic}  \&
  {Ryde}}{{Harper} et~al.}{2020}]{Harper20}
{Harper} G.~M.,  {Guinan} E.~F.,  {Wasatonic} R.,   {Ryde} N.,  2020, \mn@doi
  [\apj] {10.3847/1538-4357/abc1f0}, \href
  {https://ui.adsabs.harvard.edu/abs/2020ApJ...905...34H} {905, 34}

\bibitem[\protect\citeauthoryear{{Hillier}, {Lanz}, {Heap}, {Hubeny}, {Smith},
  {Evans}, {Lennon}  \& {Bouret}}{{Hillier} et~al.}{2003}]{Hillier03}
{Hillier} D.~J.,  {Lanz} T.,  {Heap} S.~R.,  {Hubeny} I.,  {Smith} L.~J.,
  {Evans} C.~J.,  {Lennon} D.~J.,   {Bouret} J.~C.,  2003, \mn@doi [\apj]
  {10.1086/374329}, \href
  {http://adsabs.harvard.edu/cgi-bin/nph-bib_query?bibcode=2003ApJ...588.1039H&db_key=AST}
  {588, 1039}

\bibitem[\protect\citeauthoryear{{Kravchenko} et~al.,}{{Kravchenko}
  et~al.}{2021}]{Kravchenko21}
{Kravchenko} K.,  et~al., 2021, arXiv e-prints, \href
  {https://ui.adsabs.harvard.edu/abs/2021arXiv210408105K} {p. arXiv:2104.08105}

\bibitem[\protect\citeauthoryear{{Kudritzki} \& {Puls}}{{Kudritzki} \&
  {Puls}}{2000}]{Kudritzki-Puls00}
{Kudritzki} R.-P.,  {Puls} J.,  2000, \mn@doi [\araa]
  {10.1146/annurev.astro.38.1.613}, \href
  {http://adsabs.harvard.edu/cgi-bin/nph-bib_query?bibcode=2000ARA%26A..38..613K&db_key=AST}
  {38, 613}

\bibitem[\protect\citeauthoryear{{Levesque} \& {Massey}}{{Levesque} \&
  {Massey}}{2020}]{Levesque-Massey20}
{Levesque} E.~M.,  {Massey} P.,  2020, \mn@doi [\apjl]
  {10.3847/2041-8213/ab7935}, \href
  {https://ui.adsabs.harvard.edu/abs/2020ApJ...891L..37L} {891, L37}

\bibitem[\protect\citeauthoryear{{Levesque}, {Massey}, {Olsen}, {Plez},
  {Josselin}, {Maeder}  \& {Meynet}}{{Levesque} et~al.}{2005}]{Levesque05}
{Levesque} E.~M.,  {Massey} P.,  {Olsen} K.~A.~G.,  {Plez} B.,  {Josselin} E.,
  {Maeder} A.,   {Meynet} G.,  2005, \mn@doi [\apj] {10.1086/430901}, \href
  {http://adsabs.harvard.edu/cgi-bin/nph-bib_query?bibcode=2005ApJ...628..973L&db_key=AST}
  {628, 973}

\bibitem[\protect\citeauthoryear{{Levesque}, {Massey}, {Olsen}, {Plez},
  {Meynet}  \& {Maeder}}{{Levesque} et~al.}{2006}]{Levesque06}
{Levesque} E.~M.,  {Massey} P.,  {Olsen} K.~A.~G.,  {Plez} B.,  {Meynet} G.,
  {Maeder} A.,  2006, \mn@doi [\apj] {10.1086/504417}, \href
  {http://adsabs.harvard.edu/abs/2006ApJ...645.1102L} {645, 1102}

\bibitem[\protect\citeauthoryear{{Levesque}, {Massey}, {Olsen}  \&
  {Plez}}{{Levesque} et~al.}{2007}]{Levesque07}
{Levesque} E.~M.,  {Massey} P.,  {Olsen} K.~A.~G.,   {Plez} B.,  2007, \mn@doi
  [\apj] {10.1086/520797}, \href
  {http://adsabs.harvard.edu/abs/2007ApJ...667..202L} {667, 202}

\bibitem[\protect\citeauthoryear{{Markova}, {Puls}  \& {Langer}}{{Markova}
  et~al.}{2018}]{Markova18}
{Markova} N.,  {Puls} J.,   {Langer} N.,  2018, \mn@doi [\aap]
  {10.1051/0004-6361/201731361}, \href
  {https://ui.adsabs.harvard.edu/abs/2018A&A...613A..12M} {613, A12}

\bibitem[\protect\citeauthoryear{{Massey}, {Plez}, {Levesque}, {Olsen},
  {Clayton}  \& {Josselin}}{{Massey} et~al.}{2005}]{massey05}
{Massey} P.,  {Plez} B.,  {Levesque} E.~M.,  {Olsen} K.~A.~G.,  {Clayton}
  G.~C.,   {Josselin} E.,  2005, \mn@doi [\apj] {10.1086/497065}, \href
  {http://adsabs.harvard.edu/abs/2005ApJ...634.1286M} {634, 1286}

\bibitem[\protect\citeauthoryear{{Montarg{\`e}s}, {Kervella}, {Perrin},
  {Ohnaka}, {Chiavassa}, {Ridgway}  \& {Lacour}}{{Montarg{\`e}s}
  et~al.}{2014}]{Montarges14}
{Montarg{\`e}s} M.,  {Kervella} P.,  {Perrin} G.,  {Ohnaka} K.,  {Chiavassa}
  A.,  {Ridgway} S.~T.,   {Lacour} S.,  2014, \mn@doi [\aap]
  {10.1051/0004-6361/201423538}, \href
  {https://ui.adsabs.harvard.edu/abs/2014A&A...572A..17M} {572, A17}

\bibitem[\protect\citeauthoryear{Montarg{\`e}s et~al.,}{Montarg{\`e}s
  et~al.}{2021}]{Montarges21}
Montarg{\`e}s M.,  et~al., 2021, \mn@doi [Nature] {10.1038/s41586-021-03546-8},
  594, 365

\bibitem[\protect\citeauthoryear{{Negueruela}, {Gonz{\'a}lez-Fern{\'a}ndez},
  {Dorda}, {Marco}  \& {Clark}}{{Negueruela} et~al.}{2013}]{Negueruela13}
{Negueruela} I.,  {Gonz{\'a}lez-Fern{\'a}ndez} C.,  {Dorda} R.,  {Marco} A.,
  {Clark} J.~S.,  2013, in {Kervella} P.,  {Le Bertre} T.,   {Perrin} G.,  eds,
   EAS Publications Series Vol. 60, EAS Publications Series. pp 279--285
  (\mn@eprint {arXiv} {1303.1837}), \mn@doi{10.1051/eas/1360032}

\bibitem[\protect\citeauthoryear{{Perrin}, {Ridgway}, {Coud{\'e} du Foresto},
  {Mennesson}, {Traub}  \& {Lacasse}}{{Perrin} et~al.}{2004}]{Perrin04}
{Perrin} G.,  {Ridgway} S.~T.,  {Coud{\'e} du Foresto} V.,  {Mennesson} B.,
  {Traub} W.~A.,   {Lacasse} M.~G.,  2004, \mn@doi [\aap]
  {10.1051/0004-6361:20040052}, \href
  {http://adsabs.harvard.edu/abs/2004A%26A...418..675P} {418, 675}

\bibitem[\protect\citeauthoryear{{Perrin}, {Ridgway}, {Verhoelst}, {Schuller},
  {Coud{\'e} du Foresto}, {Traub}, {Millan-Gabet}  \& {Lacasse}}{{Perrin}
  et~al.}{2005}]{Perrin05}
{Perrin} G.,  {Ridgway} S.~T.,  {Verhoelst} T.,  {Schuller} P.~A.,  {Coud{\'e}
  du Foresto} V.,  {Traub} W.~A.,  {Millan-Gabet} R.,   {Lacasse} M.~G.,  2005,
  \mn@doi [\aap] {10.1051/0004-6361:20042313}, \href
  {https://ui.adsabs.harvard.edu/abs/2005A&A...436..317P} {436, 317}

\bibitem[\protect\citeauthoryear{{Plez}}{{Plez}}{2008}]{plez08}
{Plez} B.,  2008, \mn@doi [Physica Scripta Volume T]
  {10.1088/0031-8949/2008/T133/014003}, \href
  {http://adsabs.harvard.edu/abs/2008PhST..133a4003P} {133, 014003}

\bibitem[\protect\citeauthoryear{{Plez}}{{Plez}}{2012}]{plez12}
{Plez} B.,  2012, Astrophysics Source Code Library, \href
  {http://adsabs.harvard.edu/abs/2012ascl.soft05004P} {p.~5004}

\bibitem[\protect\citeauthoryear{{Safonov} et~al.,}{{Safonov}
  et~al.}{2020}]{Safonov20}
{Safonov} B.,  et~al., 2020, arXiv e-prints, \href
  {https://ui.adsabs.harvard.edu/abs/2020arXiv200505215S} {p. arXiv:2005.05215}

\bibitem[\protect\citeauthoryear{{Smith}}{{Smith}}{2014}]{Smith14}
{Smith} N.,  2014, \mn@doi [\araa] {10.1146/annurev-astro-081913-040025}, \href
  {http://adsabs.harvard.edu/abs/2014ARA%26A..52..487S} {52, 487}

\bibitem[\protect\citeauthoryear{{Tsuji}}{{Tsuji}}{2000}]{tsuji00}
{Tsuji} T.,  2000, \mn@doi [\apj] {10.1086/309185}, \href
  {http://adsabs.harvard.edu/abs/2000ApJ...538..801T} {538, 801}

\bibitem[\protect\citeauthoryear{{Tsuji}}{{Tsuji}}{2006}]{Tsuji06}
{Tsuji} T.,  2006, \mn@doi [\apj] {10.1086/504585}, \href
  {https://ui.adsabs.harvard.edu/abs/2006ApJ...645.1448T} {645, 1448}

\bibitem[\protect\citeauthoryear{{Woitke}, {Helling}, {Hunter}, {Millard},
  {Turner}, {Worters}, {Blecic}  \& {Stock}}{{Woitke} et~al.}{2018}]{Woitke18}
{Woitke} P.,  {Helling} C.,  {Hunter} G.~H.,  {Millard} J.~D.,  {Turner} G.~E.,
   {Worters} M.,  {Blecic} J.,   {Stock} J.~W.,  2018, \mn@doi [\aap]
  {10.1051/0004-6361/201732193}, \href
  {https://ui.adsabs.harvard.edu/abs/2018A&A...614A...1W} {614, A1}

\bibitem[\protect\citeauthoryear{{Zamora}, {Garc{\'\i}a-Hern{\'a}ndez}, {Plez}
  \& {Manchado}}{{Zamora} et~al.}{2014}]{Zamora14}
{Zamora} O.,  {Garc{\'\i}a-Hern{\'a}ndez} D.~A.,  {Plez} B.,   {Manchado} A.,
  2014, \mn@doi [\aap] {10.1051/0004-6361/201423626}, \href
  {https://ui.adsabs.harvard.edu/abs/2014A&A...564L...4Z} {564, L4}

\bibitem[\protect\citeauthoryear{{van Loon}, {Groenewegen}, {de Koter},
  {Trams}, {Waters}, {Zijlstra}, {Whitelock}  \& {Loup}}{{van Loon}
  et~al.}{1999}]{vanLoon99}
{van Loon} J.~T.,  {Groenewegen} M.~A.~T.,  {de Koter} A.,  {Trams} N.~R.,
  {Waters} L.~B.~F.~M.,  {Zijlstra} A.~A.,  {Whitelock} P.~A.,   {Loup} C.,
  1999, \aap, \href
  {http://adsabs.harvard.edu/cgi-bin/nph-bib_query?bibcode=1999A%26A...351..559V&db_key=AST}
  {351, 559}

\bibitem[\protect\citeauthoryear{{van Loon}, {Cioni}, {Zijlstra}  \&
  {Loup}}{{van Loon} et~al.}{2005}]{vanLoon05}
{van Loon} J.~T.,  {Cioni} M.-R.~L.,  {Zijlstra} A.~A.,   {Loup} C.,  2005,
  \mn@doi [\aap] {10.1051/0004-6361:20042555}, \href
  {http://adsabs.harvard.edu/abs/2005A%26A...438..273V} {438, 273}

\makeatother
\end{thebibliography}



\bsp	
\label{lastpage}
\end{document}